\documentclass{article}
\usepackage{spconf,amsmath,graphicx}
\usepackage{amsfonts}
\usepackage{arydshln}
\usepackage{multirow}
\usepackage{booktabs}
\usepackage{color}
\usepackage{url}
\usepackage{float}
\usepackage{hyperref}

\title{E2 TTS: Embarrassingly Easy Fully Non-Autoregressive Zero-Shot TTS}

\name{\begin{tabular}{c}
Sefik Emre Eskimez,
Xiaofei Wang,
Manthan Thakker,
Canrun Li,
Chung-Hsien Tsai,
Zhen Xiao,\\
Hemin Yang,
Zirun Zhu,
Min Tang,
Xu Tan,
Yanqing Liu,
Sheng Zhao,
Naoyuki Kanda
\end{tabular}}
\address{Microsoft Corporation, USA}

\begin{document}
\ninept
\maketitle
\begin{abstract}
This paper introduces Embarrassingly Easy Text-to-Speech (E2 TTS), a fully non-autoregressive zero-shot text-to-speech system that offers human-level naturalness and state-of-the-art speaker similarity and intelligibility. In the E2 TTS framework, the text input is converted into a character sequence with filler tokens. The flow-matching-based mel spectrogram generator is then trained based on the audio infilling task. Unlike many previous works, it does not require additional components (e.g., duration model, grapheme-to-phoneme) or complex techniques (e.g., monotonic alignment search). Despite its simplicity, E2 TTS achieves state-of-the-art zero-shot TTS capabilities that are comparable to or surpass previous works, including Voicebox and NaturalSpeech 3. The simplicity of E2 TTS also allows for flexibility in the input representation. We propose several variants of E2 TTS to improve usability during inference. See \url{https://aka.ms/e2tts/} for demo samples. 
\end{abstract}
\section{Introduction}
\label{sec:intro}
\vspace{-.5em}

\begin{figure*}[t]
	\begin{minipage}[b]{1.0\textwidth}
		\centering
		\centerline{\includegraphics[width=\textwidth]{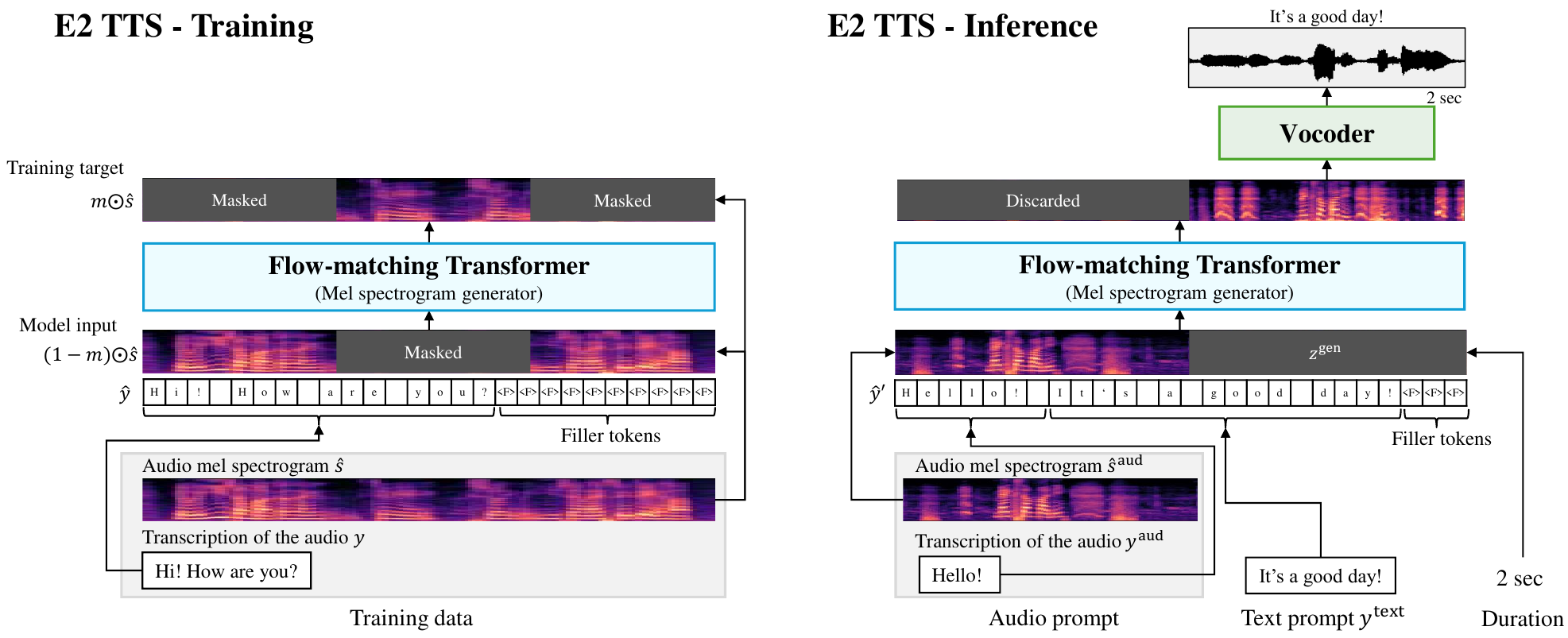}}
	\end{minipage}
 \vspace{-5mm}
	\caption{An overview of the training (left) and the inference (right) processes of E2 TTS.}
	\label{fig:main_fig}
 \vspace{-5mm}
\end{figure*}

In recent years, text-to-speech (TTS) systems have seen significant improvements \cite{ren2019fastspeech,ren2020fastspeech,kim2020glow,kong2020hifi}, achieving a level of naturalness that is indistinguishable from human speech \cite{tan2024naturalspeech}. This advancement has further led to research efforts to generate natural speech for any speaker from a short audio sample, often referred to as an audio prompt. Early studies of zero-shot TTS used speaker embedding to condition the TTS system \cite{arik2018neural,jia2018transfer}. More recently, VALL-E \cite{wang2023neural} proposed formulating the zero-shot TTS problem as a language modeling problem in the neural codec domain, achieving significantly improved speaker similarity while maintaining a simplistic model architecture. Various extensions were proposed to improve stability \cite{wang2023speechx,du2024vall,xin2024rall}, and VALL-E 2 \cite{chen2024valle} recently achieved human-level zero-shot TTS with techniques including repetition-aware sampling and grouped code modeling.

While the neural codec language model-based zero-shot TTS achieved promising results, there are still a few limitations based on its auto-regressive (AR) model-based architecture. Firstly, because the codec token needs to be sampled sequentially, it inevitably increases the inference latency. Secondly, a dedicated effort to figure out the best tokenizer (both for text tokens and audio tokens) is necessary to achieve the best quality \cite{wang2023speechx}. Thirdly, it is required to use some tricks to stably handle long sequences of audio codecs, such as the combination of AR and non-autoregressive (NAR) modeling \cite{wang2023neural,chen2024valle}, multi-scale transformer \cite{yang2023uniaudio}, grouped code modeling \cite{chen2024valle}.

Meanwhile, several fully NAR zero-shot TTS models have been proposed with promising results. Unlike AR-based models, fully NAR models enjoy fast inference based on parallel processing. NaturalSpeech 2 \cite{shen2023naturalspeech} and NaturalSpeech 3 \cite{ju2024naturalspeech} estimate the latent vectors of a neural audio codec based on diffusion models \cite{ho2020denoising,song2020score}. Voicebox \cite{le2024voicebox} and Matcha-TTS \cite{mehta2024matcha} used a flow-matching model \cite{lipman2022flow} conditioned by an input text. However, one notable challenge
for such NAR models is how to obtain the alignment between the input text and the output audio, whose length is significantly different. NaturalSpeech 2, NaturalSpeech 3, and Voicebox used a frame-wise phoneme alignment for training the model. Matcha-TTS, on the other hand, used monotonic alignment search (MAS) \cite{popov2021grad,kim2020glow} to automatically find the alignment between input and output. While MAS could alleviate the necessity of the frame-wise phoneme aligner, it still requires an independent duration model to estimate the duration of each phoneme during inference. More recently, E3 TTS \cite{gao2023e3} proposed using cross-attention from the input sequence, which required a carefully designed U-Net architecture \cite{ronneberger2015u}. 
 As such, fully NAR zero-shot TTS models require either an explicit duration model or a carefully designed architecture. One of our findings in this paper is that such techniques are not necessary to achieve high-quality zero-shot TTS, and they are sometimes even harmful to naturalness.\footnote{Concurrent with our work, Seed-TTS \cite{anastassiou2024seed} proposed a diffusion model-based zero-shot TTS, named Seed-TTS$_{\rm DiT}$. Although it appears to share many similarities with our approach, the authors did not elaborate on the details of their model, making it challenging to compare with our work.}

Another complexity in TTS systems is the choice of the text tokenizer. As discussed above, the AR-model-based system requires a careful selection of tokenizer to achieve the best result. On the other hand, most fully NAR models assume a monotonic alignment between text and output, with the exception of E3 TTS, which uses cross-attention. These models impose constraints on the input format and often require a text normalizer to avoid invalid input formats. When the model is trained based on phonemes, a grapheme-to-phoneme converter is additionally required.

In this paper, we propose \textbf{Embarrassingly Easy TTS (E2 TTS)}, a fully NAR zero-shot TTS system with a surprisingly simple architecture. E2 TTS consists of only two modules: a flow-matching-based mel spectrogram generator and a vocoder.
The text input is converted into a character sequence with filler tokens to match the length of the input character sequence and the output mel-filterbank sequence. The mel spectrogram generator, composed of a vanilla Transformer with U-Net style skip connections, is trained using a speech-infilling task \cite{le2024voicebox}.
Despite its simplicity, E2 TTS achieves state-of-the-art zero-shot TTS capabilities that are comparable to, or surpass, previous works, including Voicebox and NaturalSpeech 3.
The simplicity of E2 TTS also allows for flexibility in the input representation. We propose several variants of E2 TTS to improve usability during inference.

\section{E2 TTS}
\label{sec:method}
\vspace{-.5em}

\subsection{Training}
\vspace{-.5em}
Fig.~\ref{fig:main_fig} (a) provides an overview of E2 TTS training.
Suppose we have a training audio sample $s$ with transcription $y=(c_1, c_2, ..., c_M)$, where $c_i$ represents the $i$-th character of the transcription.\footnote{Alternatively, we can represent $y$ as a sequence of Unicode bytes~\cite{li2019bytes}.}
First, we extract its mel-filterbank features $\hat{s}\in\mathbb{R}^{D\times T}$, where $D$ denotes the feature dimension and $T$ represents the sequence length.
We then create an extended character sequence $\hat{y}$, where a special filler token $\langle F\rangle$ is appended to $y$ to make the length of $\hat{y}$ equal to $T$.\footnote{We assume $M\le T$, which is almost always valid.}
\begin{equation}
\hat{y} = (c_1, c_2, \ldots, c_M, \underbrace{\langle F\rangle, \ldots, \langle F\rangle}_{(T-M) \text{ times}}).
\end{equation}

A spectrogram generator, consisting of a vanilla Transformer~\cite{vaswani2017attention} with U-net~\cite{ronneberger2015u} style skip connection, is then trained based on the speech infilling task~\cite{le2024voicebox}.
More specifically, the model is trained
to learn the distribution $P(m\odot\hat{s}|(1-m)\odot\hat{s}, \hat{y})$, where $m\in\{0,1\}^{D\times T}$ represents a binary temporal mask, and $\odot$ 
is the Hadamard product.
E2 TTS uses the conditional flow-matching~\cite{lipman2022flow} to learn
such distribution.

\subsection{Inference}

Fig.~\ref{fig:main_fig} (b) provides an overview of the inference with E2 TTS.
Suppose we have an audio prompt $s^{aud}$ and its transcription $y^{\rm aud}=(c'_1, c'_2, ..., c'_{M^{\rm aud}})$ to mimic the speaker characteristics.
We also suppose a text prompt $y^{\rm text}=(c''_1, c''_2, ..., c''_{M^{\rm text}})$.
In the E2 TTS framework, we also require the target duration of the speech that we want to generate, which may be determined arbitrarily.
The target duration is internally represented by the frame length $T^{\rm gen}$.

First, we extract the mel-filterbank features $\hat{s}^{\rm aud}\in\mathbb{R}^{D\times T^{\rm aud}}$ from $s^{\rm aud}$.
We then create an extended character sequence $\hat{y}'$ by concatenating $y^{\rm aud}$, $y^{\rm text}$, and repeated $\langle F\rangle$, as follows:
\begin{equation}
    \hat{y}' = (c'_1, c'_2, \ldots, c'_{M^{\rm aud}}, c''_1, c''_2, \ldots, c''_{M^{\rm text}}, \underbrace{\langle F\rangle, \ldots, \langle F\rangle}_{\mathcal{T} \text{ times}}),
\end{equation}
where $\mathcal{T}=T^{\rm aud}+T^{\rm gen}-M^{\rm aud}-M^{\rm text}$, which ensures the length of $\hat{y}'$ is equal to $T^{\rm aud}+T^{\rm gen}$.\footnote{To ensure $\mathcal{T}\ge 0$, $T^{\rm gen}$ needs to satisfy $T^{\rm gen}\ge M^{\rm aud}+M^{\rm text}-T^{\rm aud}$, which is almost always valid in the TTS scenario.}

The mel spectrogram generator then generates mel-filterbank features $\tilde{s}$ based on the learned distribution of $P(\tilde{s}|[\hat{s}^{\rm aud}; z^{\rm gen}], \hat{y}')$, where $z^{\rm gen}$ is an all-zero matrix with a shape of ${D\times T^{\rm gen}}$, and $[;]$ is a concatenation operation in the dimension of $T^*$.
The generated part of $\tilde{s}$ are then converted to the speech signal based on the vocoder.

\begin{figure*}[t]
	\begin{minipage}[b]{1.0\textwidth}
		\centering
		\centerline{\includegraphics[width=\textwidth]{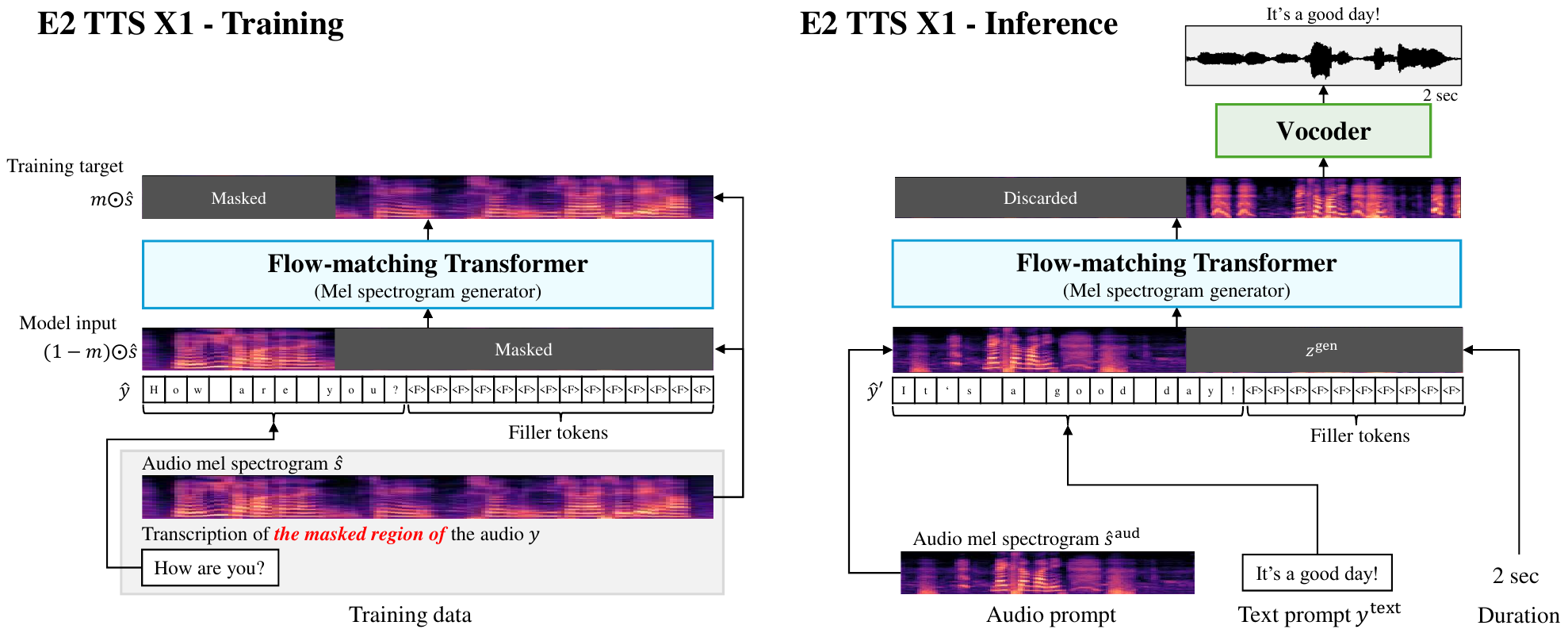}}
	\end{minipage}
 \vspace{-5mm}
	\caption{An overview of the training (left) and the inference (right) processes of E2 TTS X1.}
	\label{fig:extension_fig}
 \vspace{-5mm}
\end{figure*}

\vspace{-.5em}
\subsection{Flow-matching-based mel spectrogram generator}
\vspace{-.5em}
E2 TTS leverages conditional flow-matching~\cite{lipman2022flow}, which incorporates the principles of continuous normalizing flows~\cite{chen2018neural}. This model operates by transforming a simple initial distribution $p_0$ into a complex target distribution $p_1$ that characterizes the data. The transformation process is facilitated by a neural network, parameterized by $\theta$, which is trained to estimate a time-dependent vector field, denoted as $v_t(x;\theta)$, for $t \in [0, 1]$. From this vector field, we derive a flow, $\phi_{t}$, which effectively transitions $p_0$ into $p_1$. The neural network's training is driven by the conditional flow matching objective:
\begin{equation}
\mathcal{L}^{\text{CFM}}(\theta) = \mathbb{E}_{t,q(x_1), p_t(x|x_1)} \left\| u_t(x|x_1) - v_t(x;\theta) \right\|^2,
\end{equation}
where $p_t$ is the probability path at time $t$, $u_t$ is the designated vector field for $p_t$, $x_1$ symbolizes the random variable corresponding to the training data, and $q$ is the distribution of the training data. In the training phase, we construct both a probability path and a vector field from the training data, utilizing an optimal transport path: $p_t(x|x_1)=\mathcal{N}(x|tx_1, (1-(1-\sigma_{\rm min})t)^2I)$ and $u_t(x|x_1)=(x_1-(1-\sigma_{\rm min})x)/(1-(1-\sigma_{\rm min})t)$. For inference, we apply an ordinary differential equation (ODE)
 solver~\cite{chen2018neural} to generate the log mel-filterbank features starting from the initial distribution $p_0$.   

We adopt the same model architecture with the audio model of Voicebox (Fig. 2 of \cite{le2024voicebox})
except that the frame-wise phoneme sequence is replaced into $\hat{y}$.
Specifically,
 Transformer with U-Net style skip connection~\cite{le2024voicebox}
is used as a backbone.
The input to the mel spectrogram generator 
is $(1-m)\odot\hat{s}$, $\hat{y}$, 
the flow step $t$,
and noisy speech $s_t$.
$\hat{y}$ is first converted to character embedding 
sequence $\tilde{y}\in\mathbb{R}^{E\times T}$.
Then, $(1-m)\odot\hat{s}$, $s_t$, $\tilde{y}$ 
are all stacked to form 
a tensor with a shape of 
$(2\cdot D+E)\times T$,
followed by a linear layer to output a tensor with a shape of $D\times T$.
Finally, an embedding representation, $\hat{t}\in\mathbb{R}^D$, of $t$ is appended
to form the input tensor with a shape of $\mathbb{R}^{D\times (T+1)}$ to the Transformer.
The Transformer is trained to output a vector field $v_t$ with the conditional flow-matching objective $\mathcal{L}^{\rm CFM}$.

\vspace{-.5em}
\subsection{Relationship to Voicebox}
\vspace{-.5em}

E2 TTS has a close relationship with the Voicebox. From the perspective of the Voicebox, E2 TTS replaces a frame-wise phoneme sequence used in conditioning with a character sequence that includes a filler token. This change significantly simplifies the model by eliminating the need for a grapheme-to-phoneme converter, a phoneme aligner, and a phoneme duration model. From another viewpoint, the mel spectrogram generator of E2 TTS can be viewed as a joint model of the grapheme-to-phoneme converter, the phoneme duration model, and the audio model of the Voicebox. This joint modeling significantly improves naturalness while maintaining speaker similarity and intelligibility, as will be demonstrated in our experiments.

\begin{figure}[!t]
	\begin{minipage}[b]{1.0\columnwidth}
		\centering
		\centerline{\includegraphics[width=\columnwidth]{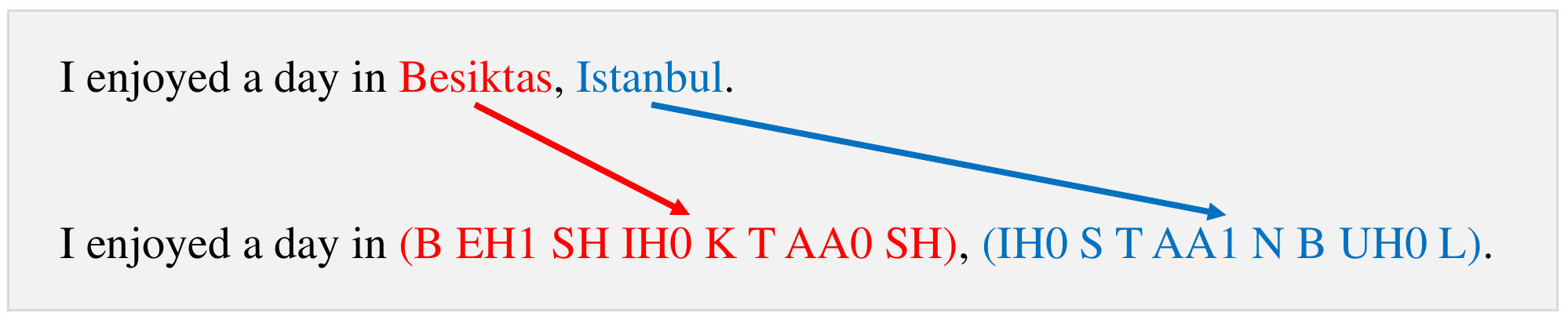}}
	\end{minipage}
 \vspace{-3mm}
	\caption{Example of the transcription for E2 TTS X2 where words are replaced with phoneme sequences enclosed in parentheses.}
	\label{fig:extension_2_fig}
\end{figure}

\vspace{-.5em}
\subsection{Extension of E2 TTS}
\vspace{-.5em}
\subsubsection{Extension 1: Eliminating the need for transcription of audio prompts in inference}
\vspace{-.5em}
\label{ssec:extension1}

In certain application contexts, obtaining the transcription of the audio prompt can be challenging. 
To eliminate the requirement of the transcription of the audio prompt
during inference,
we introduce an extension,
illustrated in Fig.~\ref{fig:extension_fig}. 
This extension, referred to as E2 TTS X1, 
assumes that we have access to the transcription of the masked region of the audio,
which we use for $y$.
During inference, the extended character sequence $\hat{y}'$ is formed 
without $y^{\rm aud}$, namely,
\begin{equation}
    \hat{y}' = (c''_1, c''_2, \ldots, c''_{M^{\rm text}}, \underbrace{\langle F\rangle, \ldots, \langle F\rangle}_{\mathcal{T} \text{ times}}).
\end{equation}
The rest of the procedure remains the same as in the basic E2 TTS.

The transcription of the masked region of the training audio can be obtained in several ways. One method is to simply apply automatic speech recognition (ASR) to the masked region during training, which is straightforward but costly. In our experiment, we employed the Montreal Forced Aligner~\cite{mcauliffe2017montreal} to determine the start and end times of words within each training data sample. The masked region was determined in such a way that we ensured not to cut the word in the middle.

\vspace{-.5em}
\subsubsection{Extension 2: Enabling explicit indication of pronunciation for parts of words in a sentence}
\vspace{-.5em}
\label{ssec:extension2}

In certain scenarios, 
users want to specify the pronunciation of a specific word
such as unique foreign names.
Retraining the model to accommodate such new words is both expensive and time-consuming.

To tackle this challenge, we introduce another extension that enables us to indicate the pronunciation of a word during inference. In this extension, referred to as E2 TTS X2, we occasionally substitute a word in $y$ with a phoneme sequence enclosed in parentheses during training, as depicted in Fig.~\ref{fig:extension_2_fig}. 
In our implementation, we replaced the word in $y$ with the phoneme sequence from the CMU pronouncing dictionary~\cite{lenzocarnegie} with a 15\% probability.
During inference, we simply replace the target word with phoneme sequences enclosed in parentheses.

It's important to note that $y$ is still a simple sequence of characters, and whether the character represents a word or a phoneme is solely determined by the existence of parentheses and their content. It's also noteworthy that punctuation marks surrounding the word are retained during replacement, which allows the model to utilize these punctuation marks even when the word is replaced with phoneme sequences.

\vspace{-.5em}
\section{Experiments}
\vspace{-.5em}
\label{sec:experiments}

\subsection{Training data}
\vspace{-.5em}
We utilized the Libriheavy dataset~\cite{kang2024libriheavy} to train our models. 
The Libriheavy dataset comprises 50,000 hours of read English speech from 6,736 speakers, accompanied by transcriptions that preserve case and punctuation marks. 
It is derived from
the Librilight~\cite{kahn2020libri} dataset contains 60,000 hours of read English speech from over 7,000 speakers. 
For E2 TTS training, we used the case and punctuated transcription
without any pre-processing.
We also used a proprietary 200,000 hours of training data to investigate the scalability of the E2 TTS model.

\vspace{-.5em}
\subsection{Model configurations}
\vspace{-.5em}

We constructed our proposed E2 TTS models using a Transformer architecture. The architecture incorporated U-Net~\cite{ronneberger2015u} style skip connections, 24 layers, 16 attention heads, an embedding dimension of 1024, a linear layer dimension of 4096. The character embedding vocabulary size was 399.\footnote{We used all characters and symbols that we found in the training data without filtering.} The total number of parameters amounted to 335 million. We modeled the 100-dimensional log mel-filterbank features, extracted every 10.7 milliseconds from audio samples with a 24 kHz sampling rate. A BigVGAN~\cite{lee2022bigvgan}-based vocoder was employed to convert the log mel-filterbank features into waveforms. The masking length was randomly determined to be between 70\% and 100\% of the log mel-filterbank feature length during training. 
In addition, we randomly dropped all the conditioning information with a 20\% probability for classifier-free guidance (CFG)~\cite{ho2022classifier}.
All models were trained for 800,000 mini-batch updates with an effective mini-batch size of 307,200 audio frames. We utilized a linear decay learning rate schedule with a peak learning rate of $7.5\times10^{-5}$ and incorporated a warm-up phase for the initial 20,000 updates. 
We discarded the training samples that exceeded 4,000 frames.

In a subset of our experiments, we initialized E2 TTS models using a pre-trained model in an unsupervised manner. This pre-training was conducted on an anonymized dataset, which consisted of 200,000 hours of unlabeled data. The pre-training protocol, which involved 800,000 mini-batch updates, followed the scheme outlined in~\cite{wang2024ISsubmit}. 

In addition, 
we trained a regression-based duration model by following 
that of Voicebox~\cite{le2024voicebox}.
It is based on a Transformer architecture, consisting of 8 layers, 8 attention heads, an embedding dimension of 512, a linear layer dimension of 2048. The training process involved 75,000 mini-batch updates with 120,000 frames. 
We used this duration model to estimate the target duration
for a fair comparison with the Voicebox baseline.
Note that we will also
show that E2 TTS is robust for different target duration
in Section \ref{sec:analysis}.

\vspace{-.5em}
\subsection{Evaluation data and metrics}
\vspace{-.5em}

In order to assess our models, we utilized the test-clean subset of the LibriSpeech-PC dataset \cite{meister2023librispeech}, which is an extension of LibriSpeech \cite{panayotov2015librispeech} that includes additional punctuation marks and casing. We specifically filtered the samples to retain only those with a duration between 4 and 10 seconds. Since LibriSpeech-PC lacks some of the utterances from LibriSpeech, the total number of samples was reduced to 1,132, sourced from 39 speakers. For the audio prompt, we extracted the last three seconds from a randomly sampled speech file from the same speaker for each evaluation sample.\footnote{The test set used in our experiments can be accessed at: \url{https://github.com/microsoft/e2tts-test-suite}.}

We carried out both objective and subjective evaluations. For the objective evaluations, we generated samples using three random seeds, computed the objective metrics for each, and then calculated their average. We computed the word error rate (WER) and speaker similarity (SIM-o).
The WER is indicative of the intelligibility of the generated samples, and for its calculation, we utilized a Hubert-large-based~\cite{hsu2021hubert} ASR system. The SIM-o represents the speaker similarity between the audio prompt and the generated sample, which is estimated by computing the cosine similarity between the speaker embeddings of both. For the calculation of SIM-o, we used a WavLM-large-based~\cite{chen2022wavlm} speaker verification model.  

For the subjective assessments, we conducted two tests: the Comparative Mean Opinion Score (CMOS) and the Speaker Similarity Mean Opinion Score (SMOS). We evaluated 39 samples for both tests, with one sample per speaker from our test-clean set. Each sample was assessed by 12 Native English evaluators. In the CMOS test~\cite{ju2024naturalspeech}, evaluators were shown the ground-truth sample and the generated sample side-by-side without knowing which was the ground-truth, and were asked to rate the naturalness on a 7-point scale (from -3 to 3), where a negative value indicates a preference for the ground-truth and a positive value indicates the opposite. 
In the SMOS test, evaluators were presented with the audio prompt and the generated sample, and asked to rate the speaker similarity on a scale of 1 (not similar at all) to 5 (identical), with increments of 1. 

\begin{table}[t]
    \caption{Objective results for LibriSpeech-PC test-clean evaluation set. 
    WER is expressed in percentage.
   $^\dagger$ Our reproduction. LL, LH, and PP stand for Librilight, Libriheavy, and Proprietary, respectively.}
    \label{tab:main_results}
    \centering
    {
    \footnotesize
    \tabcolsep = 0.5mm
        \begin{tabular}{llllcc}
            \toprule
            \textbf{ID}     & \textbf{Model}   & \textbf{Data (hours)}   & \textbf{Init} &   \textbf{WER$\downarrow$} & \textbf{SIM-o$\uparrow$}  \\ 
            \midrule
            \textbf{(GT)}   & Ground Truth         & -               & -              &   2.0                               &                 0.695                                                          \\ \midrule
            \textbf{(B1)}   & VALL-E~\cite{wang2023neural}          & LL (60K)                 &   Random              &           4.9                         &    0.500                             \\
            \textbf{(B2)}   & NaturalSpeech 3~\cite{ju2024naturalspeech}         & LL (60K)                 &      Random           &    2.6                                &             0.632          \\
            \textbf{(B3)} & Voicebox~\cite{le2024voicebox}$^\dagger$        &   LL (60K) &   Random        &       2.1                             &                  0.658             \\
            \hdashline[1pt/2pt]\hdashline[0pt/1pt]
            \textbf{(B4)} & Voicebox~\cite{le2024voicebox}$^\dagger$  & LH (50K) & Random & 2.2 & 0.667 \\
            \textbf{(B5)} & Voicebox~\cite{le2024voicebox}$^\dagger$  & LH (50K) & Pretrain~\cite{wang2024ISsubmit} & 2.2 & 0.695 \\
           \midrule 
            \textbf{(P1)}   & E2 TTS           &  LH (50K)   & Random        & 2.0 & 0.675\\
            \textbf{(P2)}   & E2 TTS           &   LH (50K)   & Pretrain~\cite{wang2024ISsubmit} & {\bf 1.9} & {\bf0.708}  \\
            \textbf{(P3)}   & E2 TTS           &   PP (200K)          & Random     &    {\bf 1.9} & 0.707      \\
            
            \bottomrule
        \end{tabular}

    }
\vspace{-3mm}
\end{table}

\begin{table}[t]
    \caption{Subjective results for LibriSpeech-PC test-clean evaluation set.
$^\dagger$ Our reproduction.}
    \label{tab:subj_results}
    \centering
    {
    \footnotesize
        \begin{tabular}{@{}lllclc@{}}
            \toprule
            \textbf{ID}    & \textbf{Model}  & \textbf{} & \textbf{CMOS$\uparrow$} & \textbf{} & \textbf{SMOS$\uparrow$} \\ \midrule 
            \textbf{(GT)} & Ground Truth &           &           0.00    &           &   $3.91_{\pm 0.13}$      \\
            \midrule
            \textbf{(B2)} & NaturalSpeech 3~\cite{ju2024naturalspeech}&           &   -0.98            &           &  \textbf{4.76}$_{\pm 0.06}$             \\
            \textbf{(B4)} & Voicebox~\cite{le2024voicebox}$^\dagger$         &           &  -0.78             &           &  4.73$_{\pm 0.06}$             \\
            \textbf{(P1)}  & E2 TTS           &           &    -0.14           &           &     $4.66_{\pm 0.07}$          \\ 
            \textbf{(P2)}  & E2 TTS           &           &    \textcolor{black}{\bf -0.05}           &           &     4.65$_{\pm 0.08}$          \\ 
            \textbf{(P3)}  & E2 TTS           &           &    \textcolor{black}{-0.18}         &           &     4.64$_{\pm 0.08}$          \\ \bottomrule
        \end{tabular}

    }

\vspace{-3mm}
\end{table}
\begin{figure*}[htb]
	\begin{minipage}[b]{1.0\columnwidth}
		\centering
		\centerline{\includegraphics[width=\columnwidth]{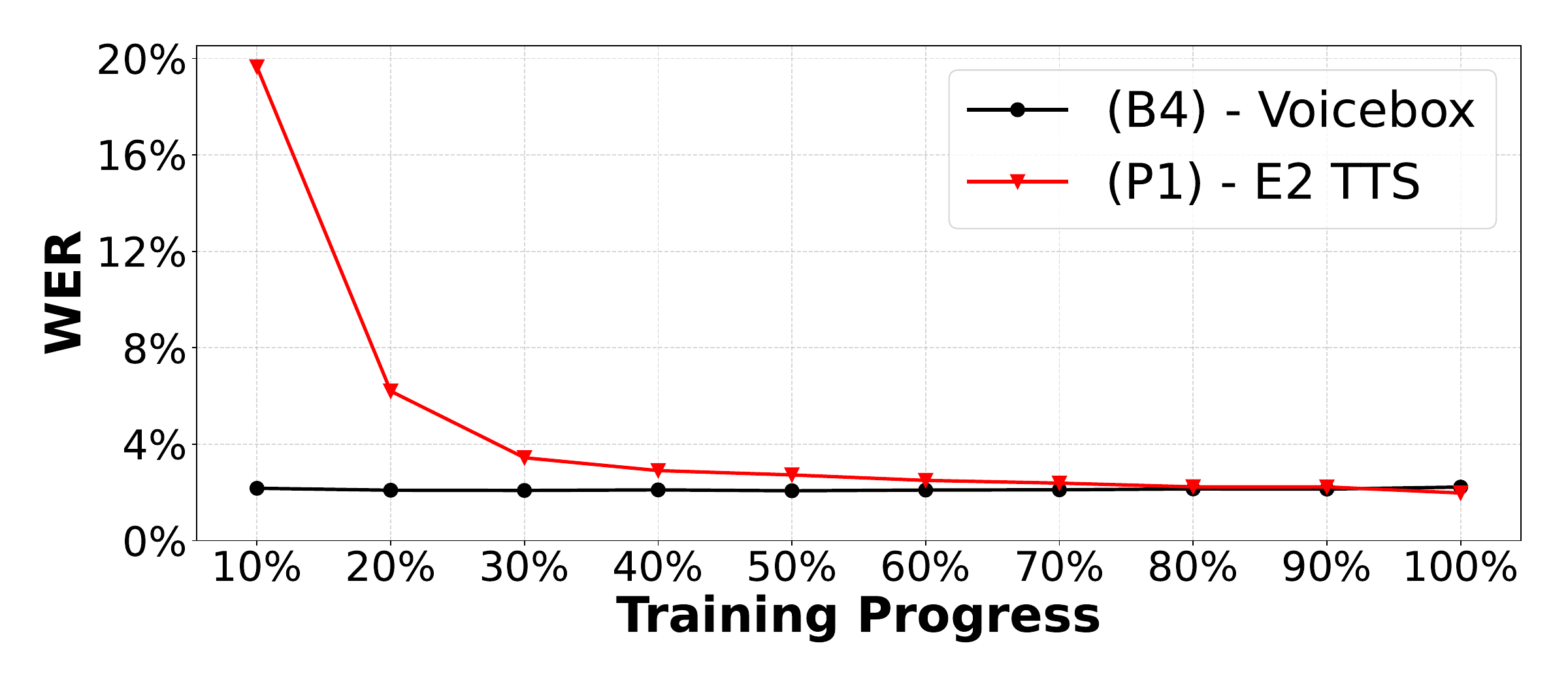}}
		  \vspace{-0.2cm}
	\end{minipage}
	\hfill
	\begin{minipage}[b]{1.0\columnwidth}
		\centering
		\centerline{\includegraphics[width=\columnwidth]{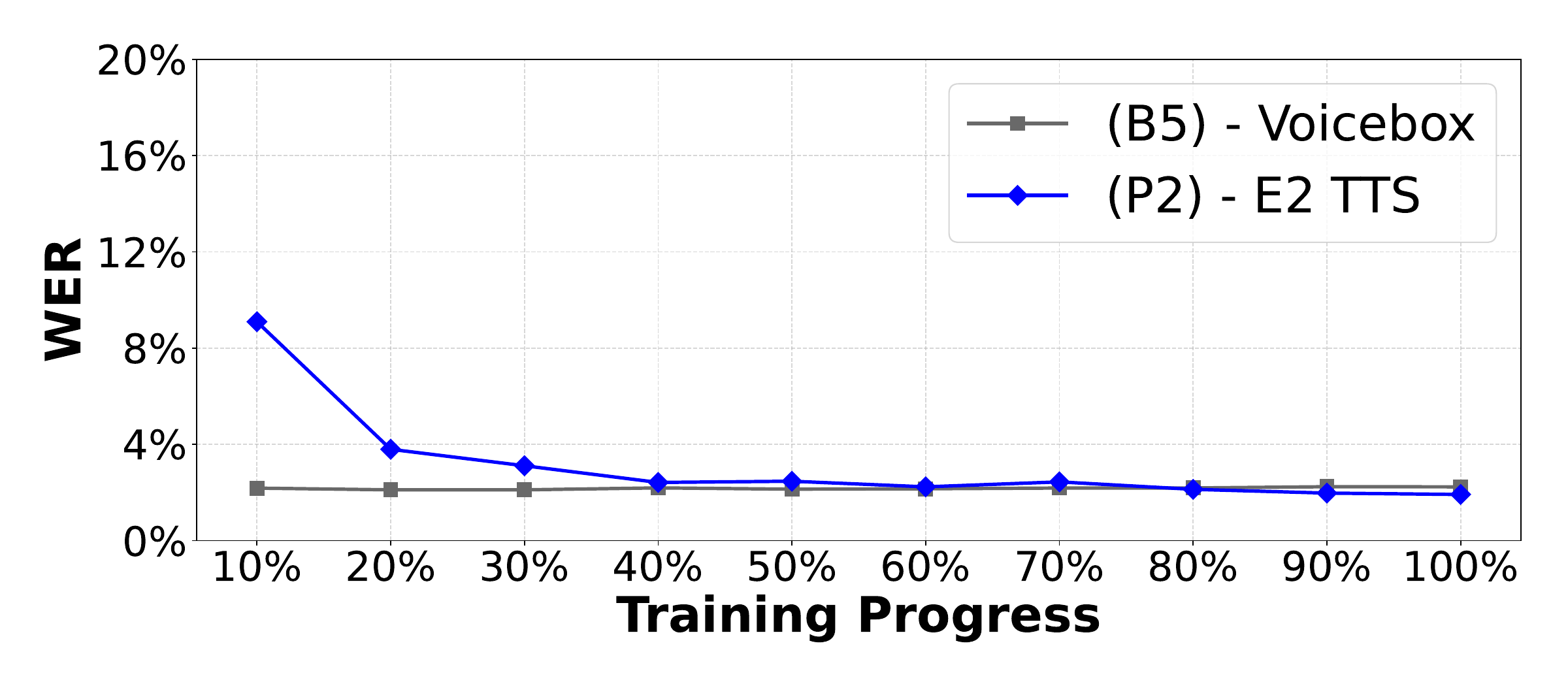}}
		  \vspace{-0.2cm}
	\end{minipage}
        \begin{minipage}[b]{1.0\columnwidth}
		\centering
		\centerline{\includegraphics[width=\columnwidth]{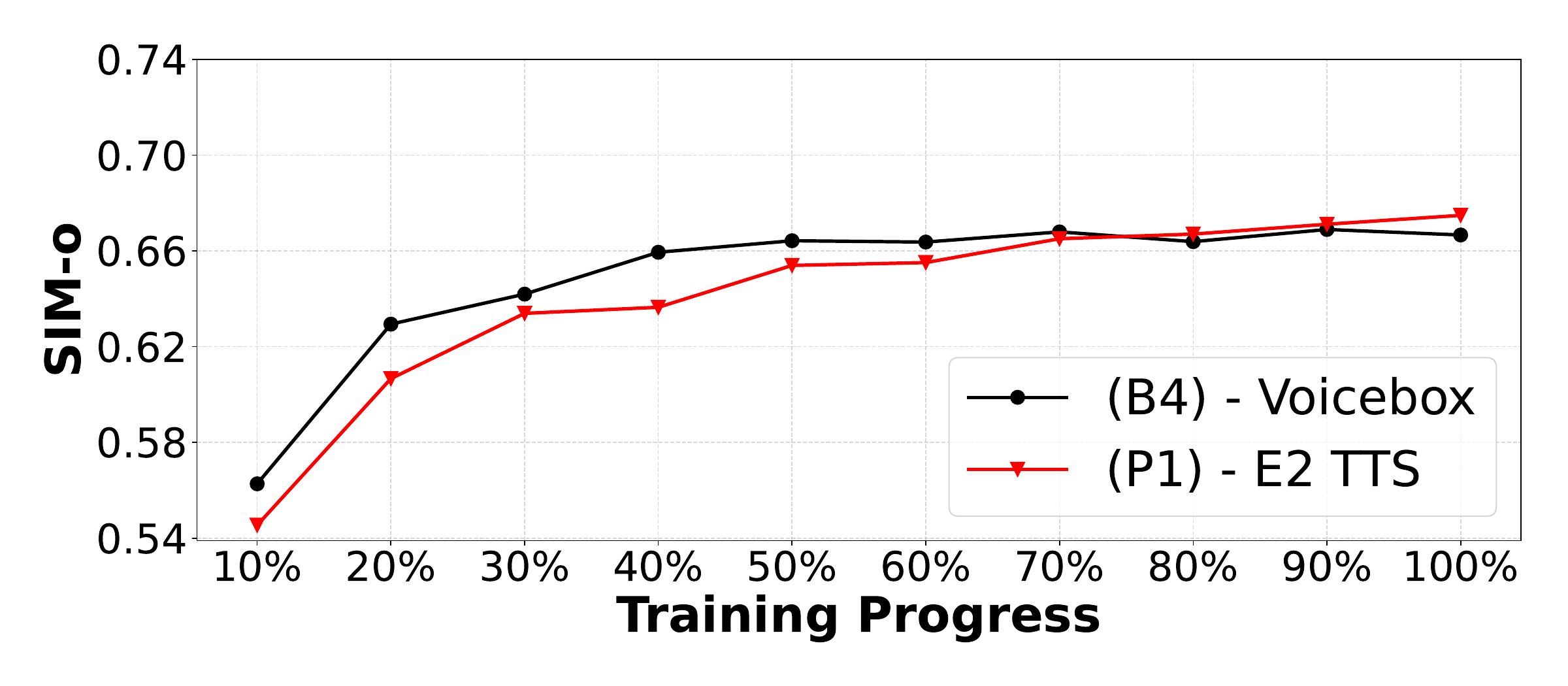}}
		\vspace{-0.2cm}
	\end{minipage}
	\hfill
	\begin{minipage}[b]{1.0\columnwidth}
		\centering
		\centerline{\includegraphics[width=\columnwidth]{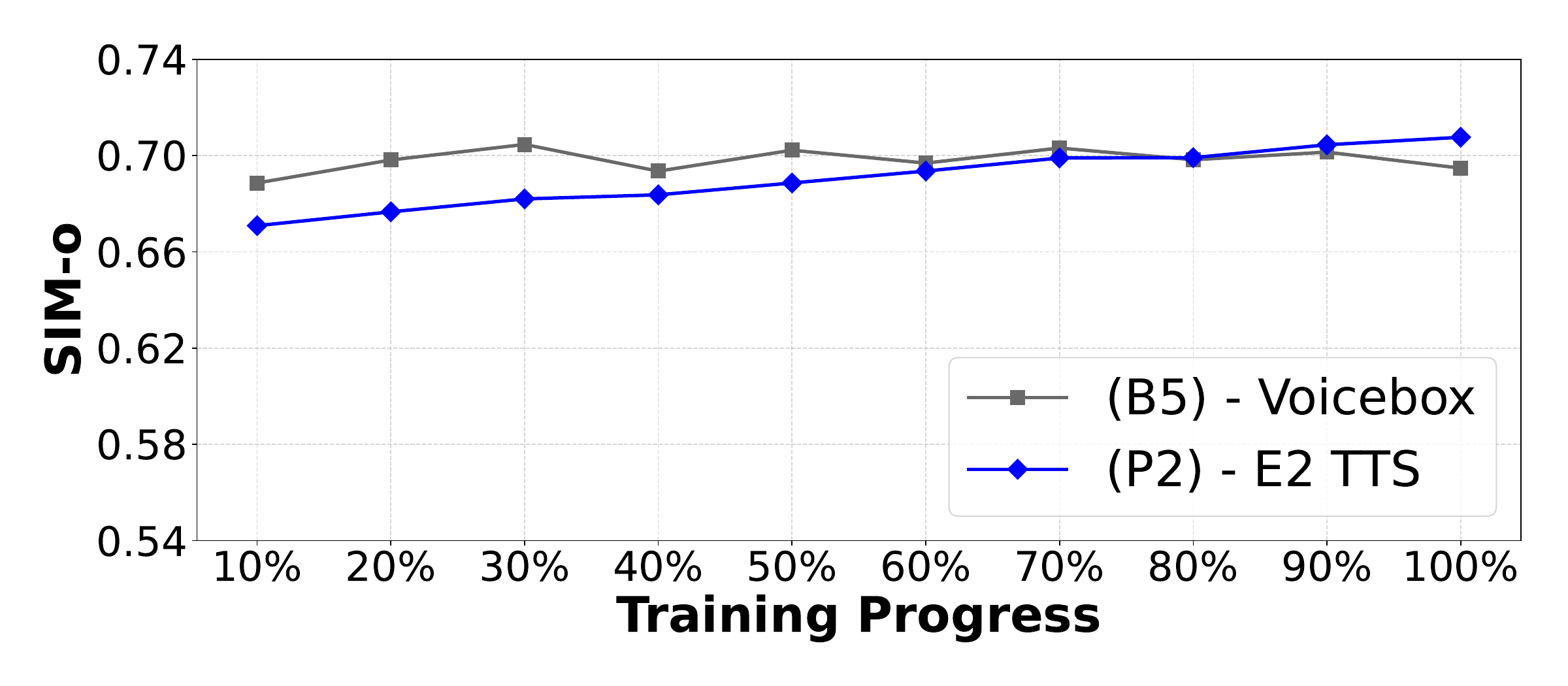}}
		\vspace{-0.2cm}
	\end{minipage}
\vspace{-3mm}
	\caption{The training progress of Voicebox and E2TTS models. The top row shows WER and the bottom row shows SIM progressions.}
	\label{fig:abl_prg_p1}
\vspace{-7mm}
\end{figure*}

\begin{table}[t]
    \caption{Comparison between the basic E2 TTS and E2 TTS X1. While the basic E2 TTS requires the transcription of the audio prompt during inference, E2 TTS X1 does not require it. 
    WER is expressed in percentage.}
    \label{tab:e2ttse1}
    \centering
    {
    \footnotesize
    \tabcolsep = 1.5mm
        \begin{tabular}{lllcc}
            \toprule
            \textbf{ID}     & \textbf{Model}     & \textbf{Init} &   \textbf{WER$\downarrow$} & \textbf{SIM-o$\uparrow$}  \\ \midrule 
            \textbf{(P1)}   & E2 TTS              & Random        &     2.0                                          &          0.675                        \\
            \textbf{(P2)}   & E2 TTS              & Pre-trained~\cite{wang2024ISsubmit}          &     1.9                             &                      0.708                 \\
            \midrule
            \textbf{(P1-X1)}   & E2 TTS X1           & Random          &   2.0                               &                 0.664                    \\
            \textbf{(P2-X1)}   & E2 TTS X1           & Pre-trained~\cite{wang2024ISsubmit}          &  2.0                                &                0.705         \\
            \bottomrule
        \end{tabular}

    }

\vspace{-5mm}
\end{table}

\vspace{-.5em}
\subsection{Main results}
\vspace{-.5em}

In our experiments, we conducted a comparison between our E2 TTS models and three other models: Voicebox~\cite{le2024voicebox}, VALL-E~\cite{wang2023neural}, and NaturalSpeech 3~\cite{ju2024naturalspeech}. We utilized our own reimplementation of the Voicebox model, which was based on the same model configuration with E2 TTS except that the Vicebox model is trained with frame-wise phoneme alignment. 
During the inference, we used CFG with a guidance strength of 1.0 for both E2 TTS and Voicebox. 
We employed the midpoint ODE solver with a number of function evaluations of 32.

Table~\ref{tab:main_results} presents the objective evaluation results for the baseline and E2 TTS models across various configurations. By comparing the (B4) and (P1) systems, we observe that the E2 TTS model achieved better WER and SIM-o than the Voicebox model when both were trained on the Libriheavy dataset. This trend holds even when we initialize the model with unsupervised pre-training~\cite{wang2024ISsubmit} ((B5) vs. (P2)), where the (P2) system achieved the best WER (1.9\%) and SIM-o (0.708) which are better than those of the ground-truth audio. Finally, by using larger training data (P3), E2 TTS achieved the same best WER (1.9\%) and the second best SIM-o (0.707) even when the model is trained from scratch, showcasing the scalability of E2 TTS. 
It is noteworthy that E2 TTS achieved superior performance compared to all strong baselines, including VALL-E, NaturalSpeech 3, and Voicebox, despite its extremely simple framework.

Table~\ref{tab:subj_results} illustrated the subjective evaluation results for NaturalSpeech 3, Voicebox, and E2 TTS. 
Firstly, all variants of E2 TTS, from (P1) to (P3), showed a better CMOS score compared to NaturalSpeech 3 and Voicebox. 
In particular, the (P2) model achieved a CMOS score of -0.05, which is considered to have a level of naturalness indistinguishable from the ground truth~\cite{ju2024naturalspeech,anastassiou2024seed}.\footnote{In the evaluation, E2 TTS was judged to be better in 33\% of samples, while the ground truth was better in another 33\% of samples. The remaining samples were judged to be of equal quality.}
The comparison between (B4) and (P1) suggests that the use of phoneme alignment was the major bottleneck in achieving better naturalness.
Regarding speaker similarity, all the models we tested showed a better SMOS compared to the ground truth, a phenomenon also observed in NaturalSpeech 3~\cite{ju2024naturalspeech}.\footnote{In LibriSpeech, some speakers utilized varying voice characteristics for different characters in the book, leading to a low SMOS for the ground truth.}
Among the tested systems, E2 TTS achieved a comparable SMOS to Voicebox and NaturalSpeech 3.

Overall, E2 TTS demonstrated a robust zero-shot TTS capability that is either superior or comparable to strong baselines, including Voicebox and NaturalSpeech 3. The comparison between Voicebox and E2 TTS revealed that the use of phoneme alignment was the primary obstacle in achieving natural-sounding audio. With a simple training scheme, E2 TTS can be easily scaled up to accommodate large training data. This resulted in human-level speaker similarity, intelligibility, and a level of naturalness that is indistinguishable from a human's voice in the zero-shot TTS setting, despite the framework’s extreme simplicity.

\begin{table}[t]
    \caption{The WER (\%) and SIM-o of E2 TTS X2 where a word is randomly replaced with the phoneme sequence during inference. Even when we replaced 50\% of words into phoneme sequences, E2 TTS X2 worked reasonably well. This indicates that we can specify the pronunciation of a new term without retraining.}
    \label{tab:e2ttse2}
    \centering
    {
    \footnotesize
    \tabcolsep = 1.2mm
        \begin{tabular}{lllccc}
            \toprule
            \textbf{ID}     & \textbf{Model}     & \textbf{Init} & \textbf{Phoneme \%} & \textbf{WER$\downarrow$} & \textbf{SIM-o$\uparrow$}  \\  
            \midrule
           \textbf{(P1)}   & E2 TTS  & Random  & 0\% & 2.0 & 0.675                                     \\
           \hdashline[1pt/2pt]\hdashline[0pt/1pt]
            \multirow{3}{*}{\textbf{(P1-X2)}}   & \multirow{3}{*}{E2 TTS X2}   & \multirow{3}{*}{Random}  & 0\%   & 2.0 &  0.679                                   \\
              &        &                           & 25\% & 2.0 & 0.678  \\ 
              &        &                           & 50\% & 2.1 & 0.679  \\
            \midrule
            \textbf{(P2)}   & E2 TTS  & Pre-trained~\cite{wang2024ISsubmit}  & 0\%  & 1.9 & 0.708                                   \\
            \hdashline[1pt/2pt]\hdashline[0pt/1pt]
            \multirow{3}{*}{\textbf{(P2-X2)}}   & \multirow{3}{*}{E2 TTS X2}   & \multirow{3}{*}{Pre-trained~\cite{wang2024ISsubmit}}  & 0\%   & 1.9 & 0.708                                    \\
              &        &                           & 25\% & 2.0 & 0.708  \\ 
              &        &                           & 50\% & 2.1& 0.707  \\ 
            \bottomrule
        \end{tabular}    
    }
\vspace{-5mm}
\end{table}

\begin{figure*}[htb]
        \hspace{-0.3cm}
        \begin{minipage}[b]{0.35\textwidth}
		\centering
		\centerline{\includegraphics[width=\columnwidth]{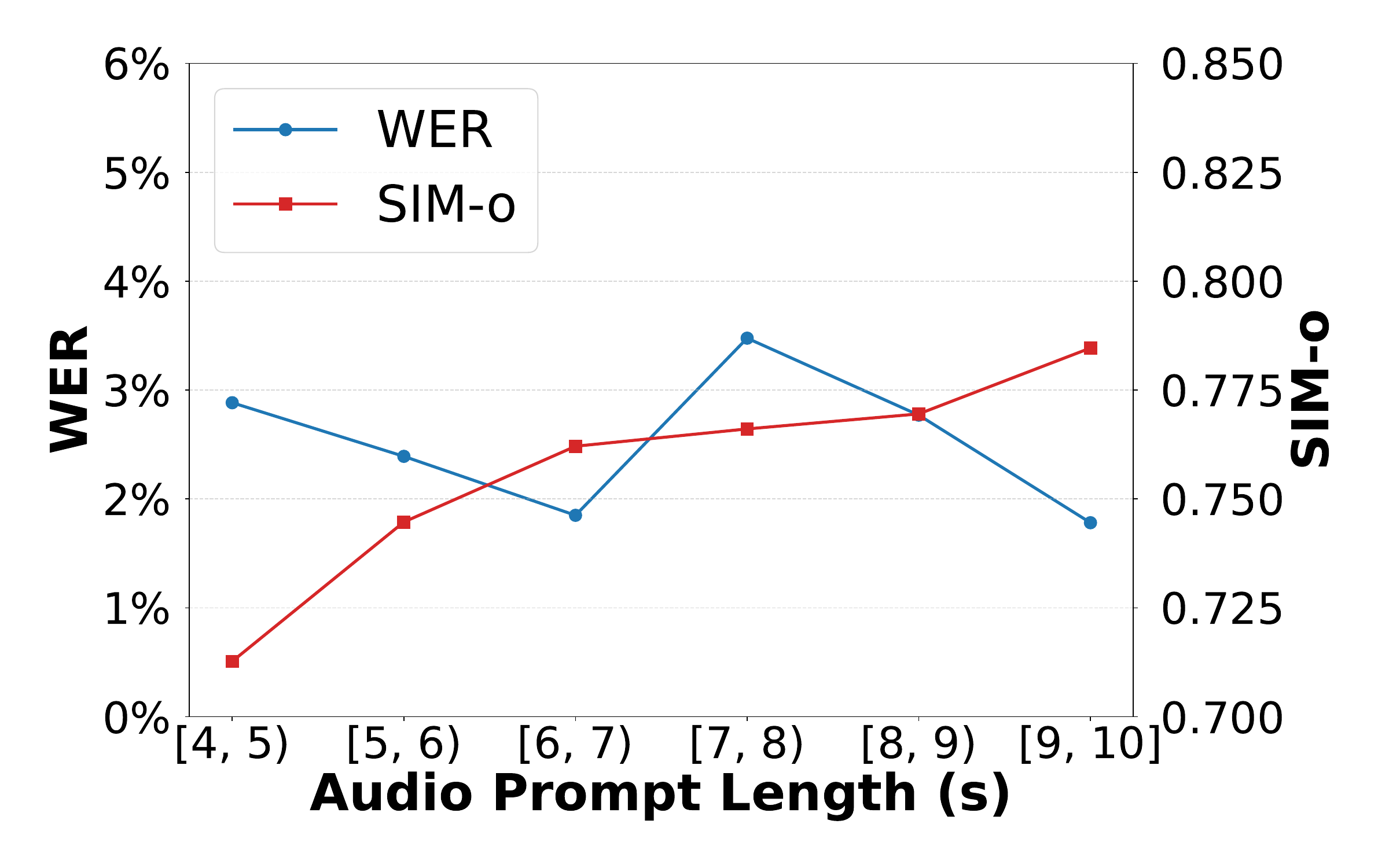}}
		  \vspace{-0.2cm}
		\centerline{(P1) E2 TTS}
	\end{minipage}
	\hspace{-0.3cm}
	\begin{minipage}[b]{0.35\textwidth}
		\centering
		\centerline{\includegraphics[width=\columnwidth]{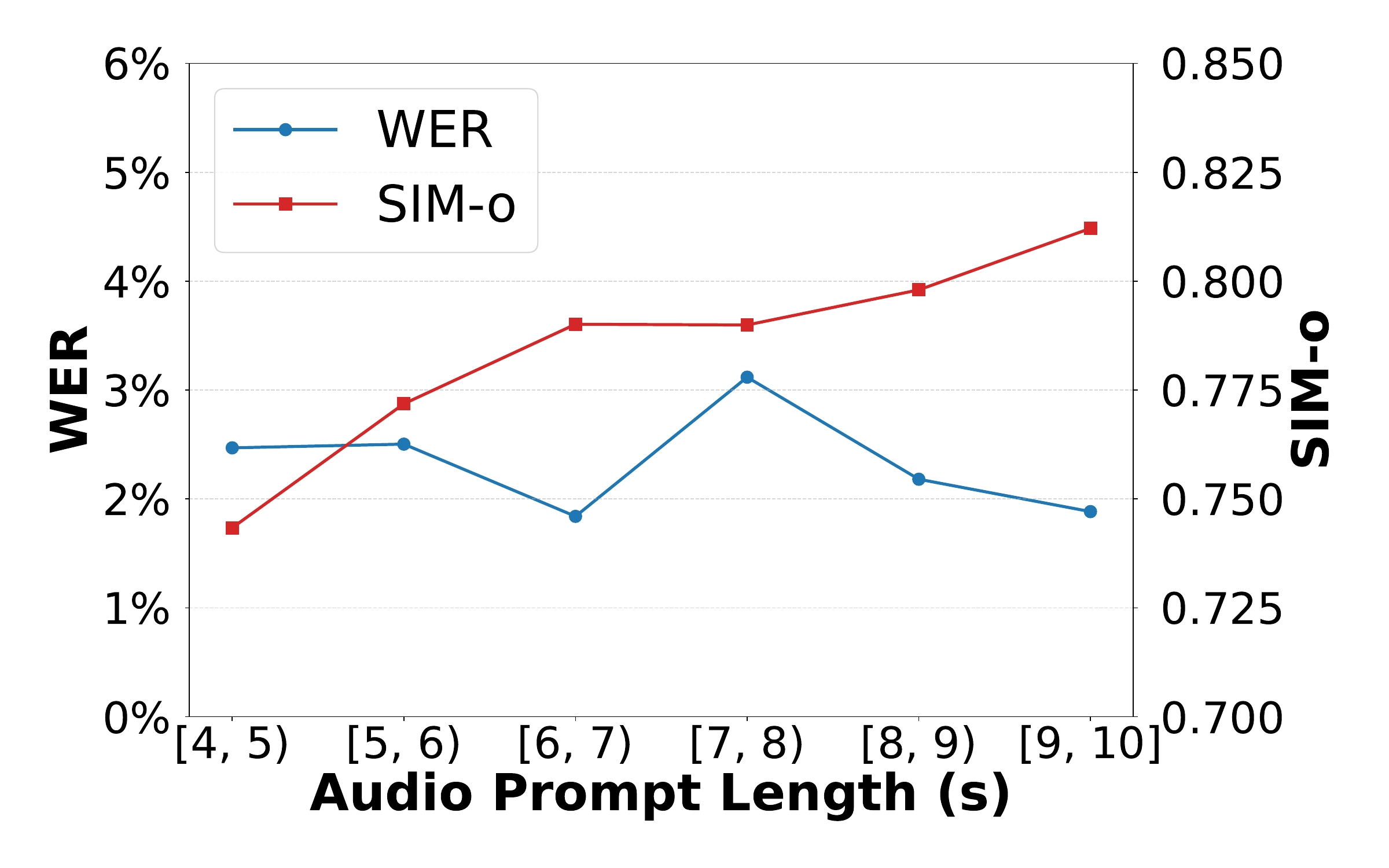}}
		\vspace{-0.2cm}
		\centerline{(P2) E2 TTS}
	\end{minipage}
        \hspace{-0.3cm}
	\begin{minipage}[b]{0.35\textwidth}
		\centering
		\centerline{\includegraphics[width=\columnwidth]{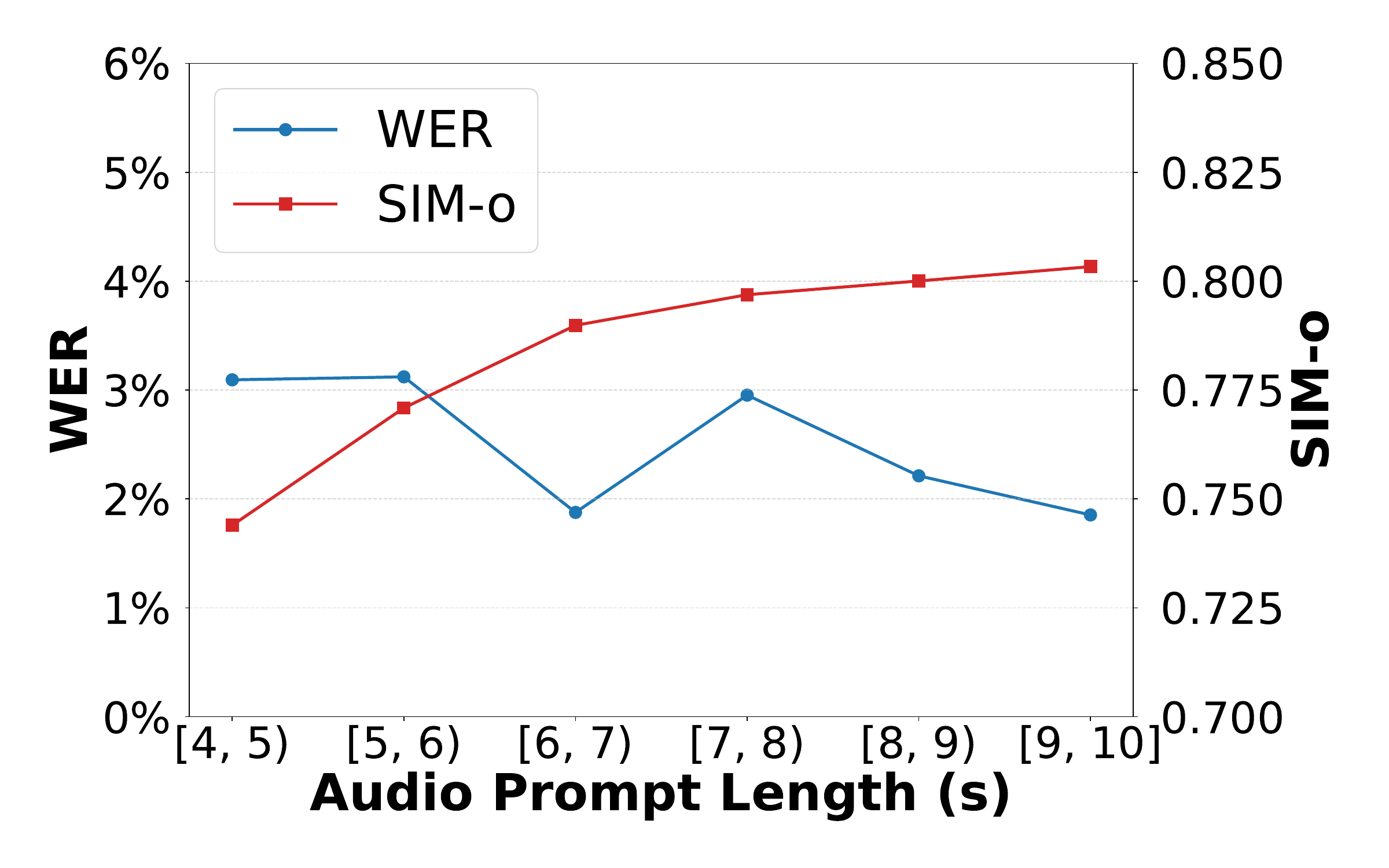}}
		\vspace{-0.2cm}
		\centerline{(P3) E2 TTS} 
	\end{minipage}
 \vspace{-5mm}
	\caption{The results of WER and SIM, shown in buckets according to the audio prompt length. The left, middle, and right plots show E2 TTS with (P1), (P2), and (P3) configurations, respectively. }
	\label{fig:abl_prompt_dur}
 \vspace{-5mm}
\end{figure*}

\vspace{-.5em}
\subsection{Evaluation of E2 TTS extensions}
\vspace{-.5em}
\subsubsection{Evaluation of the extension 1} 
The results for the E2 TTS X1 models are shown in Table~\ref{tab:e2ttse1}. These results indicate that the E2 TTS X1 model has achieved results nearly identical to those of the E2 TTS model, especially when the model was initialized by unsupervised pre-training \cite{wang2024ISsubmit}. E2 TTS X1 does not require the transcription of the audio prompt, which greatly enhances its usability.

\subsubsection{Evaluation of the extension 2} 
In this experiment, we trained the E2 TTS X2 models with a phoneme replacement rate of 15\%. During inference, we randomly replaced words in the test-clean dataset with phoneme sequences, with a probability ranging from 0\% to 50\%.

Table~\ref{tab:e2ttse2} shows the result. We first observed that E2 TTS X2 achieved parity results when no words were replaced with phoneme sequences. This shows that we can introduce extension 2 without any drawbacks. We also observed that the E2 TTS X2 model showed only marginal degradation of WER, even when we replaced 50\% of words with phoneme sequences. This result indicates that we can specify the pronunciation of a new term without retraining the E2 TTS model.     

\begin{figure}[t]
	\begin{minipage}[b]{1.0\columnwidth}
		\centering
		\centerline{\includegraphics[width=\columnwidth]{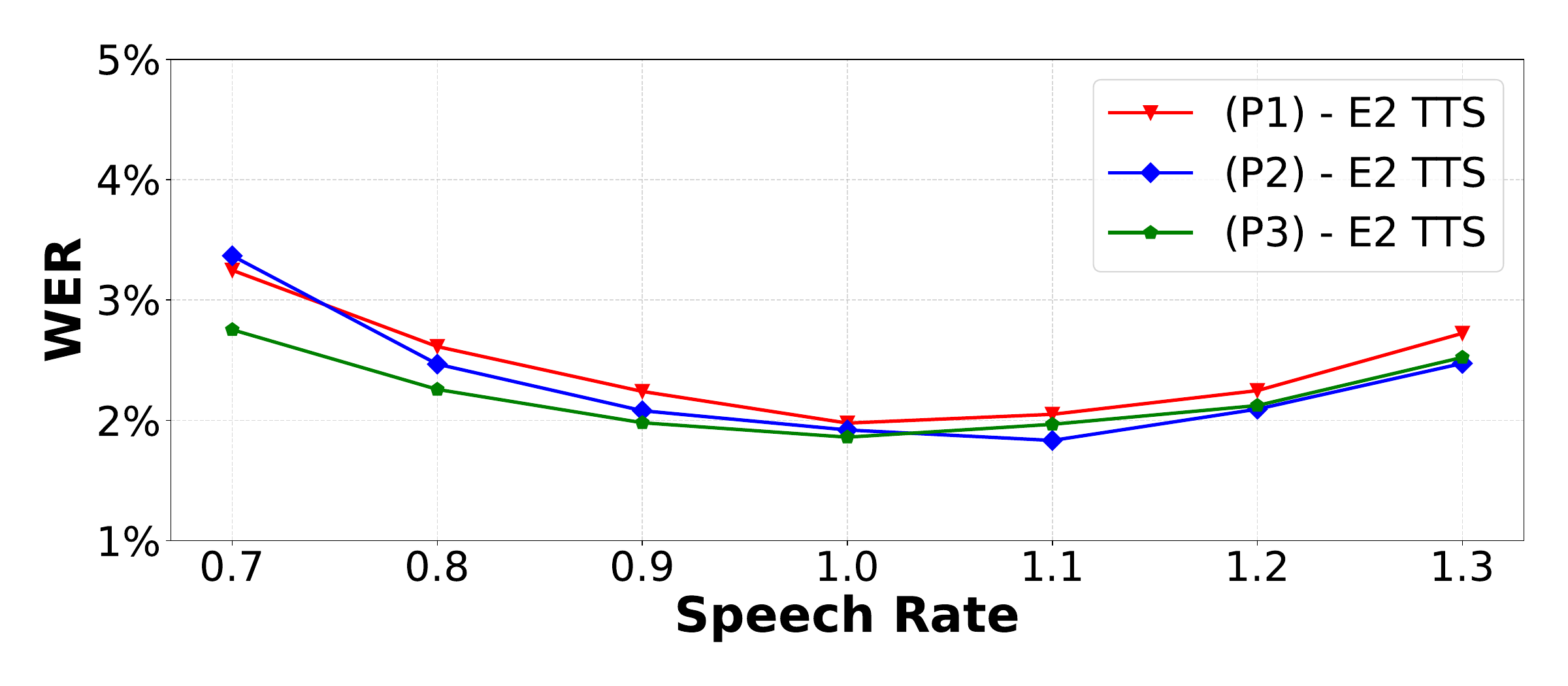}}
		  \vspace{-0.2cm}
	\end{minipage}
	\hfill
	\begin{minipage}[b]{1.0\columnwidth}
		\centering
		\centerline{\includegraphics[width=\columnwidth]{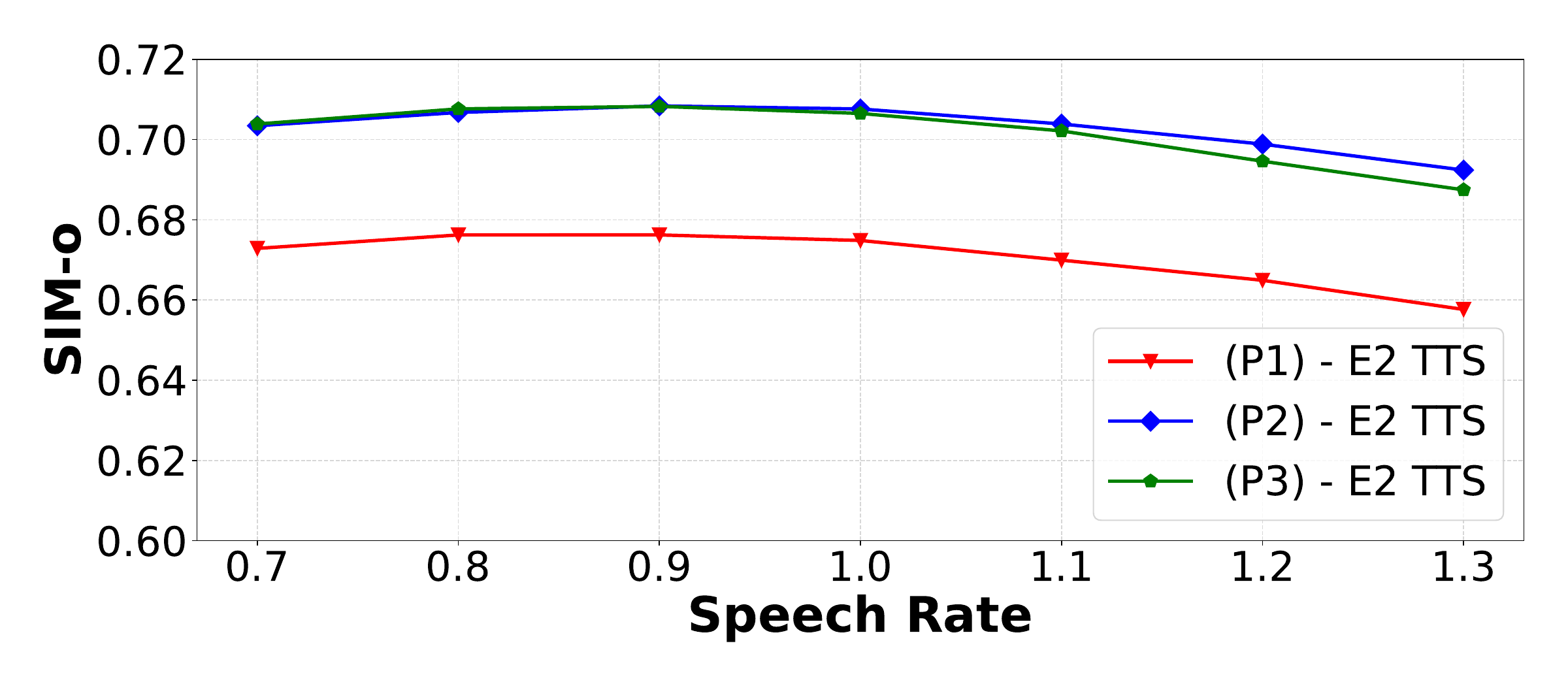}}
		\vspace{-0.2cm}
	\end{minipage}
 \vspace{-3mm}
	\caption{WER and SIM-o of different speech rates between 0.7 to 1.3 for E2TTS with (P1), (P2), and (P3) configurations.} 
	\label{fig:abl_spd}
 \vspace{-3mm}
\end{figure}

\vspace{-.5em}
\subsection{Analysis of the system behavior}
\label{sec:analysis}
\vspace{-.5em}

\subsubsection{Training progress} 
Fig.~\ref{fig:abl_prg_p1} illustrates the training progress of the (B4)-Voicebox, (B5)-Voicebox, (P1)-E2 TTS, and (P2)-E2 TTS models. The upper graphs represent the training progress as measured by WER, while the lower graphs depict the progress as measured by SIM-o.
We present a comparison between (B4)-Voicebox and (P1)-E2 TTS, as well as between (B5)-Voicebox and (P2)-E2 TTS. The former pair was trained from scratch, while the latter was initialized by unsupervised pre-training~\cite{wang2024ISsubmit}.

From the WER graphs, we observe that the Voicebox models demonstrated a good WER even at the 10\% training point, owing to the use of frame-wise phoneme alignment. On the other hand, E2 TTS required significantly more training to converge. Interestingly, E2 TTS achieved a better WER at the end of the training. We speculate this is because the E2 TTS model learned a more effective grapheme-to-phoneme mapping based on the large training data, compared to what was used for Voicebox.

From the SIM-o graphs, we also observed that E2 TTS required more training iteration, but it ultimately achieved a better result at the end of the training. We believe this suggests the superiority of E2 TTS, where the audio model and duration model are jointly learned as a single flow-matching Transformer. 

\subsubsection{Impact of audio prompt length} 

During the inference of E2 TTS, the model needs to automatically identify the boundary of $y^{\rm aud}$ and $y^{\rm text}$ in $\hat{y}$ based on the audio prompt $\hat{s}^{\rm aud}$. Otherwise, the generated audio may either contain a part of $y^{\rm aud}$ or miss a part of $y^{\rm text}$. This is not a trivial problem, and E2 TTS could show a deteriorated result when the length of the audio prompt $\hat{s}^{\rm aud}$ is long.

To examine the capability of E2 TTS, we evaluated the WER and SIM-o with different audio prompt lengths. The result is shown in Fig.~\ref{fig:abl_prompt_dur}. In this experiment, we utilized the entire audio prompts instead of using the last 3 seconds, and categorized the result based on the length of the audio prompts.
As shown in the figure, we did not observe any obvious pattern between the WER and the length of the audio prompt. This suggests that E2 TTS works reasonably well even when the prompt length is as long as 10 seconds. On the other hand, SIM-o significantly improved when the audio prompt length increased, which suggests the scalability of E2 TTS with respect to the audio prompt length.

\subsubsection{Impact of changing the speech rate} 
We further examined the model's ability to produce suitable content when altering the total duration input. 
In this experiment, we adjusted the total duration by multiplying it by $\frac{1}{sr}$, where $sr$ represents the speech rate. The results are shown in Fig.~\ref{fig:abl_spd}.
As depicted in the graphs, the E2 TTS model exhibited only a moderate increase in WER while maintaining a high SIM-o, even in the challenging cases of $sr=0.7$ and $sr=1.3$. This result suggests the robustness of E2 TTS with respect to the total duration input.

\vspace{-.5em}
\section{Conclusions}
\vspace{-.5em}
\label{sec:conclusions}
We introduced E2 TTS, a novel fully NAR zero-shot TTS.
In the E2 TTS framework,
the text input is
converted into a character sequence with filler tokens to match the
length of the input character sequence and the output mel-filterbank
sequence.
The flow-matching-based mel spectrogram generator is then trained
based on the audio infilling task.
 Despite its simplicity, E2 TTS achieved state-of-the-art zero-shot TTS capabilities that were comparable to or surpass previous works, including Voicebox and NaturalSpeech 3. 
  The simplicity of E2 TTS
also allowed for flexibility in the input representation. We proposed
several variants of E2 TTS to improve usability during inference.

\bibliographystyle{IEEEbib}
\bibliography{strings,refs}

\end{document}